\documentstyle[12pt,epsf]{article}

\begin{document}
\parskip=0.3cm
\begin{titlepage}

\hfill \vbox{\hbox{DFPD 99/TH/31}\hbox{UNICAL-TH 2/99}\hbox{BITP-99-3E}
\hbox{July 1999}}

\vskip 0.3cm

\centerline{\bf TRIPLE POMERON AND PROTON DIFFRACTION DISSOCIATION $~^\diamond$}

\vskip 0.7cm

\centerline{R.~Fiore$^{a\dagger}$, A.~Flachi$^{b\ddagger}$,
L.L.~Jenkovszky$^{c\S}$, F.~Paccanoni$^{d\ast}$, A.~Papa$^{a\dagger}$}

\vskip .3cm

\centerline{$^{a}$ \sl  Dipartimento di Fisica, Universit\`a della Calabria,}
\centerline{\sl Istituto Nazionale di Fisica Nucleare, Gruppo collegato di Cosenza}
\centerline{\sl Arcavacata di Rende, I-87036 Cosenza, Italy}
\vskip .3cm
\centerline{$^{b}$ \sl Physics Department, University of Newcastle upon Tyne,} 
\centerline{\sl Newcastle upon Tyne, NE1 7RU, United Kingdom}
\vskip .3cm
\centerline{$^{c}$ \sl  Bogoliubov Institute for Theoretical Physics,}
\centerline{\sl Academy of Sciences of the Ukrain}
\centerline{\sl 252143 Kiev, Ukrain}
\vskip .3cm
\centerline{$^{d}$ \sl  Dipartimento di Fisica, Universit\`a di Padova,}
\centerline{\sl Istituto Nazionale di Fisica Nucleare, Sezione di Padova}
\centerline{\sl via F. Marzolo 8, I-35131 Padova, Italy}

\vskip 0.3cm

\begin{abstract}
We consider proton diffraction dissociation in the dipole
Pomeron model, where the Pomeron is represented by a double
pole in the $J-$plane, and show that unitarity can be satisfied
without decoupling of the triple Pomeron vertex. 
Differential and total diffractive cross sections for the reaction
$\bar{p}+p \to \bar{p}+X$ are analyzed and reproduced in this model.

\end{abstract}

\vskip .3cm

\hrule

\vskip.1cm

\noindent
$^{\ast}${\it Work supported by the Ministero italiano
dell'Universit\`a e della Ricerca Scientifica e Tecnologica and by the
INTAS.}
\vfill
$
\begin{array}{ll}
^{\dagger}\mbox{{\it e-mail address:}} &
   \mbox{FIORE,PAPA~@CS.INFN.IT} \\
^{\ddagger}\mbox{{\it e-mail address:}} &
   \mbox{ANTONINO.FLACHI~@NEWCASTLE.AC.UK} \\
^{\S}\mbox{{\it e-mail address:}} &
 \mbox{JENK~@BITP.KIEV.UA} \\
^{\ast}\mbox{{\it e-mail address:}} &
   \mbox{PACCANONI~@PD.INFN.IT}
\end{array}
$

\vfill
\end{titlepage}
\eject
\textheight 210mm
\topmargin 2mm
\baselineskip=24pt

\section{Introduction}

Diffractive high energy elastic scattering for hadrons appears to find
a satisfactory explanation in the framework of the Regge theory with the
exchange of the Pomeron trajectory. Rising cross sections can be
accounted for by assuming a Pomeron intercept slightly higher than one~\cite{DL} 
or, in a QCD approach, by considering the Pomeron as a
gluon ladder~\cite{LIP}. The growth of total cross sections can also be
described in a way compatible with the Froissart bound in the eikonal
model~\cite{CY} or by assuming that the Pomeron is represented by a
double pole in the complex $J-$plane~\cite{BJ}.

In contrast with the above picture, inclusive diffractive collisions in
proton-proton or antiproton-proton scattering, where one of the initial
particles changes only slightly its momentum and appears in the final
state isolated in rapidity, seem to require deep modifications to the
standard Regge models. 
    The basic problem with diffraction dissociation, known for long~\cite{BW}, 
is that the integrated cross section $\sigma_{SD}$ appears to grow faster than the
total cross section $\sigma_T$, thus violating unitarity. For example, in 
the case of a supercritical Pomeron with
$\alpha(0)=1+\delta$, $\sigma_{SD}$ grows twice as fast, $\sim 
s^{2\delta}$,
as the total cross section does, $\sigma_T\sim s^{\delta}$. The only way to 
resolve this discrepancy seemed to require the vanishing of the 
triple Pomeron coupling (Pomeron decoupling~\cite{BW}), which however 
contradicts the experimental data~\cite{GL2}.

A number of different unitarization recipes have been proposed in order
to modify the energy dependence of the predicted cross section. Eikonal
corrections~\cite{GLM} succeed in reproducing the main features of
single diffraction at high energy, while the same effect can be reached
by the inclusion of cuts in the Regge theory~\cite{KPT}. Recently a
different, more phenomenological, approach has been 
considered~\cite{GM,GL1,SCH}. Renormalization~\cite{GM,GL1} or damping~\cite{SCH}
of the Pomeron flux, that consists in setting a limit to the probability
that the proton emits a Pomeron, allow for a rising of the total
diffraction cross section compatible with experimental data. All the
above approaches are based on a supercritical Pomeron input with a
Pomeron intercept larger than one.

Apart from the incompatibility with the experimentally rising cross
sections, a unity intercept Pomeron would present analogous problems
with unitarity~\cite{BW}. If, however, the partial wave amplitude, for the
Pomeron exchange, presents a simple and a double pole in the complex
$J-$plane, cross sections will grow with energy and it will be possible to
satisfy unitarity at the Born level, without eikonalization~\cite{BJ}.
In all (or most) of the models explored until now, the Pomeron was 
assumed to be an isolated single Regge pole. From QCD we know however~\cite{LIP} 
that the Pomeron is not a single pole, but rather a complicated set of 
singularities in the $J-$plane. A simple and feasible way to approximate this
complicated structure is to take the sum of a simple and a double pole (dipole).
The dipole Pomeron is known~\cite{BJ} to have unique properties since 
it reproduces itself under unitarization and thus one expects that it can
be used also to resolve the abovementioned problem in diffraction 
dissociation.
     Obviously, the sum of a simple and double pole - like any combination 
of Regge singularities - looses factorizability, although each term remains
factorizable. Since Regge pole factorization appears to be in conflict 
with experimental results~\cite{GM}, the approach we consider is 
favoured.

The dipole Pomeron model has been tested successfully in elastic hadron-hadron
and $\gamma-$hadron reactions~\cite{BJ,MB,PD1,PD2,PD3} and an
application to single diffractive dissociation has been
considered in Refs.~\cite{JMP,FJP}. It turns out that, in this approach,
the Pomeron contribution consists of two terms, one increasing like the
logarithm of the energy and the other being energy independent,
multiplied by "a priori" different $t-$dependent vertex functions. This
feature, and the assumption that the Pomeron couples in a different way
to Pomerons and hadrons, opens the way to a unified treatment of elastic
and production amplitudes.

Since the inclusive process of hadron diffraction has been discussed
extensively in the literature~\cite{BW,RR,PDB,ABK,AG,GL2} we will start,
in Section 2, from the Mueller discontinuity formula and adapt it to the
chosen model. The triple Pomeron contribution will be discussed in
detail and the possibility to satisfy the unitarity constraint will be
investigated. While a proof of the proposed solution cannot be given in
the framework of the Regge theory because the $t-$dependence of the vertices is
arbitrary to a great extent, plausibility arguments can be advanced on
the basis of dynamical models for the Pomeron. Section 3 will be devoted 
to the inclusion of secondary Regge
trajectories and the final expression for the cross section will be
compared with experimental data in Section 4. The conclusion of
this work will be drawn in Section 5.

\section{The triple Pomeron in diffractive dissociation}

Consider first the process $a+b\to c+X$ with the exchange of
Regge trajectories $\{i\}$. From the Mueller discontinuity formula~\cite{MUE} we get
\begin{equation}
\pi E_c\frac{d^3\sigma}{d\vec{p}_c}=
\frac{1}{16\pi s}\sum_X\left|\sum_i\beta^i_{a\bar{c}}(t)
\xi_i(t)F^{ib\to X}(M^2,t)\left(\frac{s}{M^2}\right)^{\alpha_i
(t)}\right|^2
\label{z1}
\end{equation}
in the usual Regge pole model. $M^2$ is the squared mass of the
unrevealed state $X$, $\alpha_i(t)$ represents the Regge trajectory
exchanged and
\begin{displaymath}
\xi_i(t)=\frac{1\pm\exp(-i\pi\alpha_i(t))}{\sin(\pi\alpha_i(t))}
\end{displaymath}
is its signature.
In the following $i=P, f,\pi$ and $\omega$, where $P$ stands for
the Pomeron trajectory.

Consider now the elastic scattering and suppose that, asymptotically, the
absorptive part in the $s$-channel, $A(s,t)$, goes like
\begin{displaymath}
A(s,t) \propto \beta_1(t) \beta_2(t) s^{\alpha(t)} [h(t) \ln s +C]~,
\end{displaymath}
then the partial wave amplitude presents a simple and double pole in
the complex $J-$plane. The amplitude for the Pomeron exchange can then
be written as
\begin{equation}
T(s,t) \propto -\frac{(-is)^{\alpha(t)}}{\sin(\pi
\alpha(t)/2)}\beta_1(t) \beta_2(t) \left[h(t) \left(\ln s -i\frac{\pi}{2}\right)
+C\right]~,
\label{z2}
\end{equation}
where constant terms have been collected in $C$.
The explicit form of $h(t)$ depends on the model. As an example, in a
dual model, if the residue of the simple pole has the form
$\beta(\alpha(t))$, the residue of the double pole will be given by
$\int\beta(\alpha)\,d\alpha+const$~\cite{BJ}.
The form of this residue is such that the coefficient of the double pole
can vanish for $t=0$, if this is required from general principles.

To substantiate this possibility, we can generalize the picture of Ref.~\cite{LN} 
and suppose that the Pomeron pole couples to quarks through
the exchange of two gluons. In order to describe the gluon-Pomeron-gluon
vertex we use, at high energy, the rules of covariant Reggeization~\cite{GJ,TO} 
for the coupling of the Pomeron pole to massive vector mesons,
since the gluons are off-shell. This require the introduction of five
unknown functions of $t$ (some of them will vanish because of gauge
invariance) that will appear in the amplitude multiplied by polynomials
in $\alpha(t)$ and by appropriate powers of $\nu$, $\nu^{\alpha(t)-n}$,
where $\nu \approx s/2$ at high energy and $n=0,1,\ldots $.
The final expression is complicated, but it can be easily seen that there
will appear, among the leading contributions $\nu^{\alpha(t)}$,
at least one term vanishing when $\alpha(t)=1$. This
property remains true when we integrate over the gluon and quark
momenta and, by taking the derivative with respect to $\alpha(t)$, we
introduce a double pole for the Pomeron.  Moreover this
result does not depend on the choice of the partonic wave function of
the hadron. The conclusion is that, also in elastic hadron-hadron
scattering, the presence of a term vanishing with $t$, together with
other terms finite at $t=0$, is highly probable in the conventional Regge residue. 
It could well happen that, if we consider the triple Pomeron
vertex, the unitarity condition will impose constraints on the
couplings such that $h(t)$ in Eq.~(\ref{z2}) vanishes at $t=0$.

In the dipole Pomeron approach, Eq.~(\ref{z1}) becomes
\begin{eqnarray}
& &  \frac{d^2\sigma}{dM^2\,dt}=
\nonumber \\
& &  \frac{1}{16\pi s^2}\sum_X\left| \beta^{P}_{a\bar{c}}(t)
\left(-i\frac{s}{M^2}\right)^{\alpha_{P}(t)}\left[h(t)\left(\ln
\frac{s}{M^2}-i\frac{\pi}{2}\right)+C\right]F^{P b\to X}(M^2,t) \right.
\nonumber \\
& & + \left. \sum_{i\neq P} \beta^i_{a\bar{c}}\xi_i(t) F^{ib\to X}(M^2,t)
\left(\frac{s}{M^2}\right)^{\alpha_i(t)} \right|^2~.
\label{z3}
\end{eqnarray}
Let us consider now the triple Pomeron contribution to Eq.~(\ref{z3}),
neglecting for the moment all the interference terms and replacing the
sum over intermediate states by a discontinuity in $M^2$,
\begin{eqnarray}
& &\frac{1}{16\pi s^2}[\beta^{P}_{a\bar{c}}(t)]^2\left(\frac{s}{M^2}
\right)^{2\alpha_{P}(t)}\times
\nonumber \\
 & &  \left[\left(h(t)\ln\frac{s}{M^2}+C\right)^2+\frac{\pi^2}{4}h^2(t)\right]
Im\,T^{P b}(M^2,t,\alpha_{P}(t),t_{b\bar{b}}=0)~,
\label{z4}
\end{eqnarray}
where, according to Eq.~(\ref{z2}),
\begin{equation}
Im\,T^{P b}=\sigma_0\,(M^2)^{\alpha_{P}(0)}(\lambda+
\bar{h}(0)\ln M^2 +\lambda' (M^2)^{\alpha_f(0)-1}) g(t)~,
\label{z5}
\end{equation}
$g(t)$ being the triple Pomeron coupling. A term, decreasing with
$M^2$, is present in Eq.~(\ref{z5}) since we consider also the secondary $f$
trajectory in $P-b$ scattering. Obviously, if $h(0)$ vanishes, the same
will be true for $\bar{h}(0)$. In the following,
$\alpha_{P}(t)=1+\alpha't$, $\alpha'=0.25$ GeV$^{-2}$ and the standard
form for the residue will be assumed: $\beta^P_{a\bar{c}}=\exp(bt)$.

By integrating Eq.~(\ref{z5}) over $t$ and $M^2$ we get the
Pomeron contribution to the single diffractive cross-section,
$\sigma_{SD}$.
We will now show  that the constraint $\sigma_{SD}<\sigma_T$ for all
values
of $s$ requires that $h(t)\propto (-t)^{\gamma}$ with $\gamma > 1/2$.
Without changing the asymptotic behaviour of the Pomeron-hadron vertex,
we can assume that\footnote{Other choices are possible as well: the 
function $h(t) \propto 1-\exp(\bar{\gamma} t)$ has a similar 
behaviour for suitable values of $\bar{\gamma}$.}
\begin{equation}
h(t) \propto \left(\frac{-t}{-t+1} \right)^{\gamma}~,  \mbox{~~~~~~}
\gamma \geq 0~.
\label{h(t)}
\end{equation}
The proportionality constant in the expression for $h(t)$ is unessential 
since it can be factorized out in (\ref{z4}) by properly  rescaling the
constant $C$.

The proof becomes simpler if, according to experimental findings~\cite{GM,GL2}, 
we consider the triple Pomeron vertex $g(t)$ as constant
and neglect $Im T^{Pb}$ in Eq.~(\ref{z4}). Then, setting $B=2 (b+\alpha'
\ln(s/M^2))$ and $y=\ln(s/M^2)$, the $t$ integral can be easily
evaluated and reads 
\begin{displaymath}
\int_0^{\infty}\,dt\,e^{-Bt} \left[\left(\left(\frac{t}{t+1}
\right)^{\gamma}+C\right)^2+\frac{\pi^2}{4}\left(\frac{t}{t+1}
\right)^{2 \gamma} \right] =
\end{displaymath}
\begin{equation}
\Gamma(2\gamma+1) \Psi (2\gamma +1,2;B) \left(y^2+\frac{\pi^2}{4}
\right) +2 \, y \, C \, \Gamma(\gamma +1) \, \Psi(\gamma 
+1,2;B)+\frac{C^2}{B}~,
\label{z6}
\end{equation}
where $\Psi(a,c;x)$ is a confluent hypergeometric function~\cite{BAT}.

In order to integrate over $M^2$ we transform to the variable $B$ whose
upper limit is, asymptotically, proportional to $\ln s$.
The integral can be evaluated
exactly by using the elementary relations for the $\Psi$ function and,
in the limit $B\sim \ln s \to \infty$, the behaviour of $\sigma_{SD}$
can be inferred from the large variable estimate for $\Psi$~\cite{BAT}
\begin{equation}
\sigma_{SD} \: \sim \: \Gamma(2 \gamma+1)\, \frac{2}{1-\gamma} \: B^{2-2\gamma}+
\ldots + C^2 \ln B ~,
\label{z7}
\end{equation}
where dots in the l.h.s. stand for terms with a less singular behaviour 
when $s \to \infty$.

We note that the singularity for $\gamma=1$ in Eq.~(\ref{z7})
is spurious; the exact result does not present singularities for
$\gamma \geq 0$.
Since, in the model considered, $\sigma_T \sim \ln s$ and $\sigma_{SD} <
\sigma_T$, from the first term in Eq.~(\ref{z7}) we must have $2-2\gamma
\leq 1$.
Hence, the parameter $\gamma$, in general, must satisfy the condition 
$\gamma\geq 1/2$.
This inequality is necessary to avoid terms, violating unitarity,
that rise faster than $\ln s$.
It is important to notice that the triple Pomeron
contribution does not vanish at $t=0$ because of the presence of the
constant $C$.

\section{Non-leading contributions and the differential cross section}

From now on we select the hadrons participating the process: $a$ and $c$
are antiprotons ($\bar p$) and $b$ is a proton ($p$). Later, for the 
evaluation of the
total single diffractive cross section $\sigma_{SD}$, the process
$a=c=p$ and $b=\bar{p}$ will be also taken into account.

On the basis of historical fits~\cite{RR,FF}, the $\omega$ trajectory
can be neglected and, since the $\pi$ trajectory contributes in a
different kinematical region with respect to $P$ and $f$, interference
terms between $\pi$ and $P, f$ are suppressed. Hence, in Eq.~(\ref{z3})
the sum over $i$ refers only to $f$, and the $\pi$ contribution will be
chosen as in~\cite{GL2,FF,MBH,UA8}
\begin{equation}
\left. \frac{d^2\sigma}{dM^2\,dt}\right|_{\pi} \: = \:
\frac{1}{4\pi}\frac{g^2_{\pi pp}}{4\pi}\frac{(-t)}{(t-\mu^2)^2}
\left(\frac{s}{M^2}\right)^{2\alpha_{\pi}(t)-1}G^2(t)
\sigma^{\pi p}_T(M^2)~,
\label{z8}
\end{equation}
where
\begin{displaymath}
G(t)=\frac{2.3-\mu^2}{2.3-t}~,
\end{displaymath}
$g^2_{\pi pp}/(4 \pi)=14.6$ and $\alpha_{\pi}(t)=0.9\,t$.

The $f$ contribution, and its interference with the Pomeron, must now be
considered. The approximation suggested in~\cite{DL2,DL3} is based on
the assumption that the $f$ couples to hadrons in just the same way as the
Pomeron. This choice avoids the proliferation of free parameters and is
justified from the consideration that, while the $f$ is required by the
data~\cite{GLM,UA8,DL2,DL3}, its contribution is small, in percentage, and
can be approximated. A Pomeron-Pomeron-Reggeon term larger than 
$0.15 \: \sigma_{SD}$ is
excluded by high energy data~\cite{CDF} and is completely ignored in a
recent analysis~\cite{GM}.

Since the model we consider for the Pomeron is different from the
conventional, supercritical one, we must take care in choosing an
appropriate $f$ trajectory. Fits with a double Pomeron pole~\cite{PD1,PD2,PD3} 
require an intercept $\alpha_f(0)$ higher than the
value, usually adopted, about $0.55$~\cite{DL}. In a recent analysis~\cite{CKK}, 
however, for the degenerate $a_2/f$ trajectory, the result
$\alpha_+-1=-0.31\pm 0.05$ has been obtained by refitting
all the experimental cross
sections considered in~\cite{DL}. For all data, with errors added in
quadrature, a smaller value for $\alpha_+$ has been obtained: $\alpha_+
-1=-0.34 \pm 0.05$. An intermediate value, $-0.32$ has been used in Ref.~\cite{UA8}.
The coincidence of $\alpha_f(0)-1$ with $-0.32$
obtained in the fit of hadronic cross sections within different models
for the soft Pomeron should not be surprising; in a limited energy range a behaviour
$s^{\epsilon}$, for $\epsilon$ sufficiently small, can be well
approximated by a term of the form $(u+v \ln s)$.

Let  $a(t)$ be the difference between the $P$ and $f$ trajectories. If we set  
\begin{displaymath}
a(t)=\alpha_{P}(t)-\alpha_f(t)= a(0)-\delta t~,
\end{displaymath}
then typical values, adopted in the following, are $a(0)\simeq 0.34$
and $\delta\simeq 0.65$.
The $f$ contribution
\begin{eqnarray}
R(s,t) &=& k\left\{\left[h(t)\ln\frac{s}{M^2}+C\right] \cos\left(\frac{\pi a(t)}{2}
\right)- \right.
\nonumber \\
 & &  \left. \frac{\pi h(t)}{2}\sin\left(\frac{\pi
a(t)}{2}\right)\right\} \left(\frac{s}{M^2}
\right)^{-a(t)}+k^2\left(\frac{s}{M^2}\right)^{-2a(t)}
\label{z9}
\end{eqnarray}
will appear in the final form of the differential cross section:
\begin{eqnarray}
\frac{d^2\sigma}{dt\,dM^2}  &=&  \frac{A}{M^2}
e^{2(b+\alpha'\ln(s/M^2))t}
\left[\left(h(t) \ln\frac{s}{M^2}+ C\right)^2 \right. \nonumber \\
&+ & \left. \frac{\pi^2}{4} h^2(t)+
 R(s,t) \right] (1+l (M^2)^{\alpha_f(0)-1})
\nonumber  \\
&+ &  \frac{1}{4\pi}\frac{g^2}{4\pi M^2}\frac{(-t)}{(t-\mu^2)^2}G^2(t)
\left(\frac{s}{M^2}\right)^{2\alpha_{\pi}(t)-2} \sigma^{\pi p}_T
(M^2)~,
\label{z10}
\end{eqnarray}
where all the constant factors have been collected in $A$.

In Ref.~\cite{DL2} a value near $7.8$ is quoted for the parameter $k$, 
appearing in $R(s,t)$; since, however, the expression (\ref{z10}) has been
rescaled, $k$ is here a new parameter.
As far as the other parameters are concerned, $b$ will be fixed
from $p-p$ elastic scattering (e.g. $b=2.25$ GeV$^{-2}$, 
consistent with the slopes used in~\cite{GM,MB,PD1,PD2,PD3,
FJP}) and $\sigma^{\pi p}_T(M^2)$ in the dipole Pomeron model can
be written as
\begin{equation}
\sigma^{\pi p}_T(M^2)=0.565+2.902 \ln(M^2)+44.388 (M^2)^{\alpha_f(0)-1}~,
\label{z11}
\end{equation}
inspired by the parametrization used in~\cite{DL}.
Since the form of $h(t)$ is determined only near $t=0$, it is well
possible that the $t$-dependence of the cross-section should
be corrected. Hence, a different value of $b$ could be required
from the experimental data, but this possibility will not be considered
in the following.

\begin{figure}[htb]
\begin{center}
{\parbox[t]{5cm}{\epsfysize 12cm \epsffile{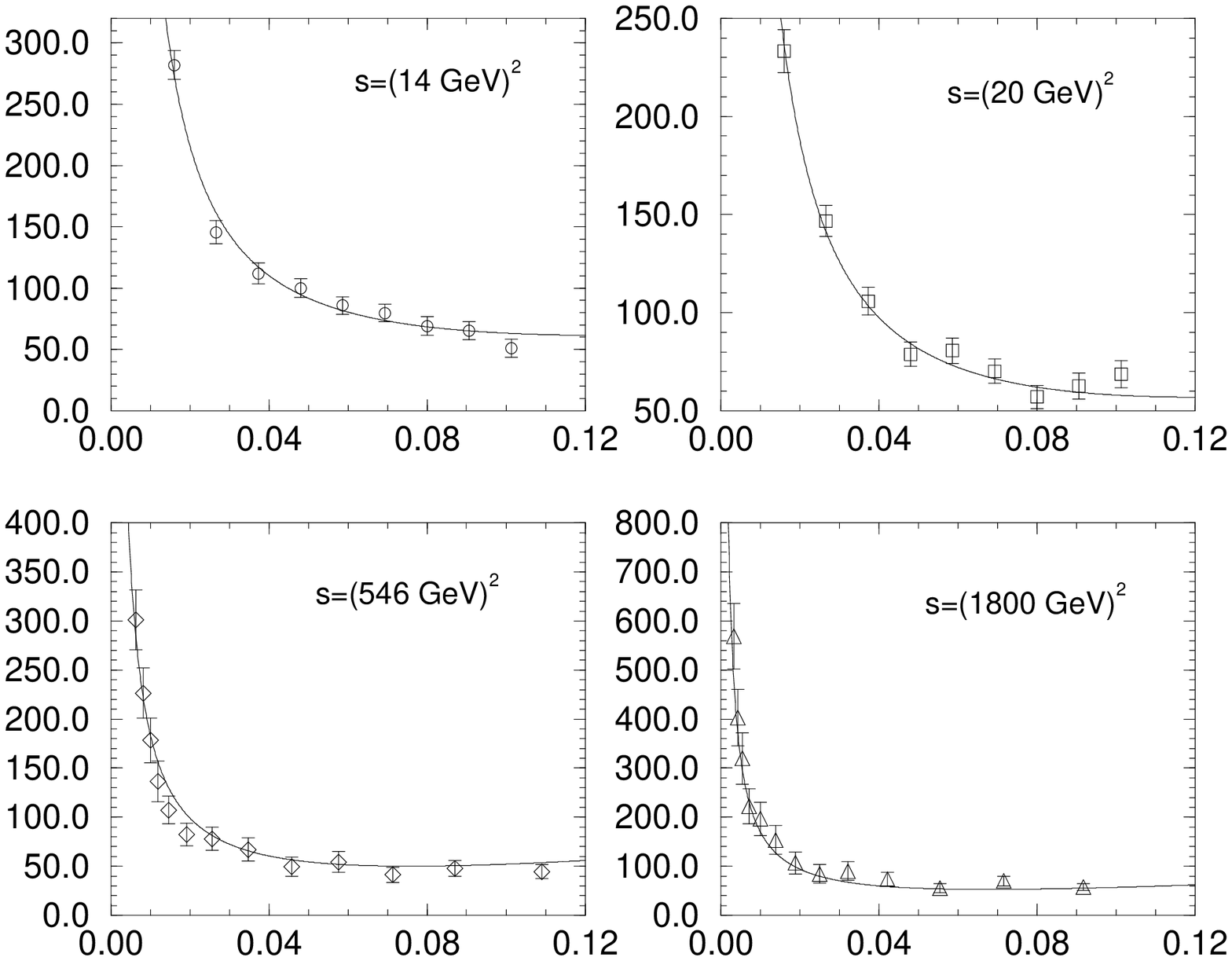}}}
\end{center}
\caption[]{Differential cross sections $d^2\sigma/d\xi\,dt$ [mb/GeV$^2$]
{\it vs} $\xi$ for several values of $s$. Data are from E396~\cite{COOL} and
CDF~\cite{CDF}, compiled in~\cite{GM}. The solid curves represent the model 
with $\gamma = 2$ and with the remaining parameters determined by the fit.}
\end{figure}

\section{Comparison with data}

When comparing the model with experimental data, we find two kinds
of problems. The first one is related to the experimental definition
of single diffraction dissociation. The great variety of
phenomenological models, adopted by different experimental groups
in order to extract the published data, makes the test of any
new model difficult. Moreover, integrated cross-sections do not refer to
the same intervals of $M^2$ and $t$, for different experimental
analyses.
The second kind of problem resides in our parametrization and is
strongly related to the first one. The integrated cross section cannot 
be given in compact form and, since the overall normalization of the 
data has an experimental uncertainty of 15 \%, it is not an easy task to 
determine the parameter $\gamma$ only from the $t$-dependence of the 
cross sections at different energies.

While the pion contribution
can be fixed as in Section 3, the parameters relative to the $f$
trajectory are different with respect to those of Refs.~\cite{DL2,DL3}, 
since the Pomeron
contribution differs from the one proposed there.
We are left with three parameters for the Pomeron and one for the $f$,
plus an overall constant multiplying these contributions, while the
$\pi$ term has no free parameters.

From now on, we adopt the standard variable $\xi\equiv M^2/s$, that 
represents the fraction of the momentum of the proton carried by the
Pomeron.
Using the expression~(\ref{z10}) for $d^2\sigma/d\xi dt$ in our model, 
we performed a global fit of the data at $\sqrt s = 14$ and 20 GeV of 
E396~\cite{COOL} and at $\sqrt s = 546$ and 1800 GeV of the 
CDF collaboration~\cite{CDF}. All the data
were taken from the compilation of Ref.~\cite{GM} and are at fixed 
$t=-0.05$ GeV$^2$.
The range of $\xi$ for the data of E396 has been 
limited to $0.0160 \: \div \: 0.1013$;
in the case of the CDF data, we have considered $\xi$ in the range 
$0.0064 \: \div \: 0.109$
for the data at $\sqrt s = 546$ GeV and in the range $0.0033 \: \div \: 0.0918$ 
for those at $\sqrt s = 1800$ GeV. We have found that our proposed model nicely fits
all the data for a large range of values of the parameter $\gamma$ larger
than 1/2. This weak dependence on the value of the parameter $\gamma$ was 
not unexpected, since the fit was performed at fixed $t$. In the particular
case of $\gamma=2$ (which will be justified in the following) the 
fit gives for the remaining parameters the following values: 
$C=0.9802$, $A=1.9080$, $k=0.9839$ and $l=2.3987$, with $\chi^2/\mbox{d.o.f.}
 \approx 0.9$.
In Fig.~1 we compare the curve resulting from the fit with $\gamma=2$ with 
the experimental data. We can 
see that our model succeeds in reproducing the experimental data at different
values of $s$. We have checked that choosing a different value for $\gamma$ 
produces only little changes of the other parameters, but does not 
affect in a sizeable way the shape of the fitting curves. 

\begin{figure}[htb]
\begin{center}
{\parbox[t]{5cm}{\epsfysize 12cm \epsffile{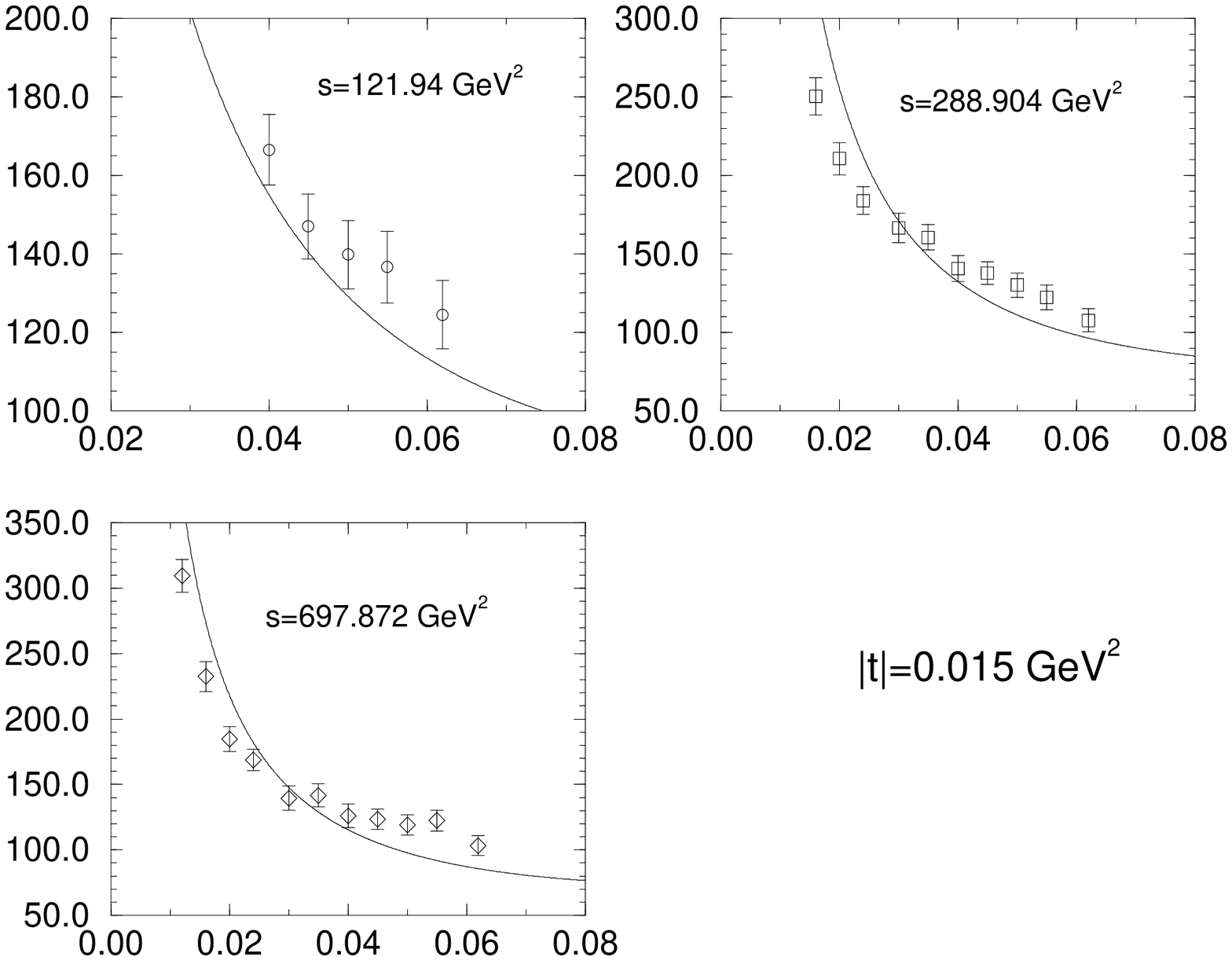}}}
\end{center}
\caption[]{Differential cross sections $d^2\sigma/d\xi\,dt$ [mb/GeV$^2$]
{\it vs} $\xi$ for several values of $s$ and $t=-0.015$ GeV$^2$. The 
solid curves represent the model with $\gamma = 2$ and with the other
parameters determined by the fit shown in Fig.~1. Data are 
from~\cite{akim77} and were not included in the fit.}
\end{figure}

We have then fixed the parameters in the expression for $d^2\sigma/d\xi dt$ 
in our model according to the result of the fit at $t=-0.05$ GeV$^2$ and 
have checked how it reproduces
other sets of data, obtained at different $t$ values. We have considered 
the data of Ref.~\cite{akim77} at $t=-0.015$ GeV$^2$ and those
of the UA8 collaboration~\cite{UA8} at the relatively large value of 
$t=-0.95$ GeV$^2$. In both cases our curves roughly reproduce the data
(see Figs.~2 and 3), thus indicating that also the $t-$dependence in our model 
is quite reasonable.

\begin{figure}[htb]
\begin{center}
{\parbox[t]{5cm}{\epsfysize 12cm \epsffile{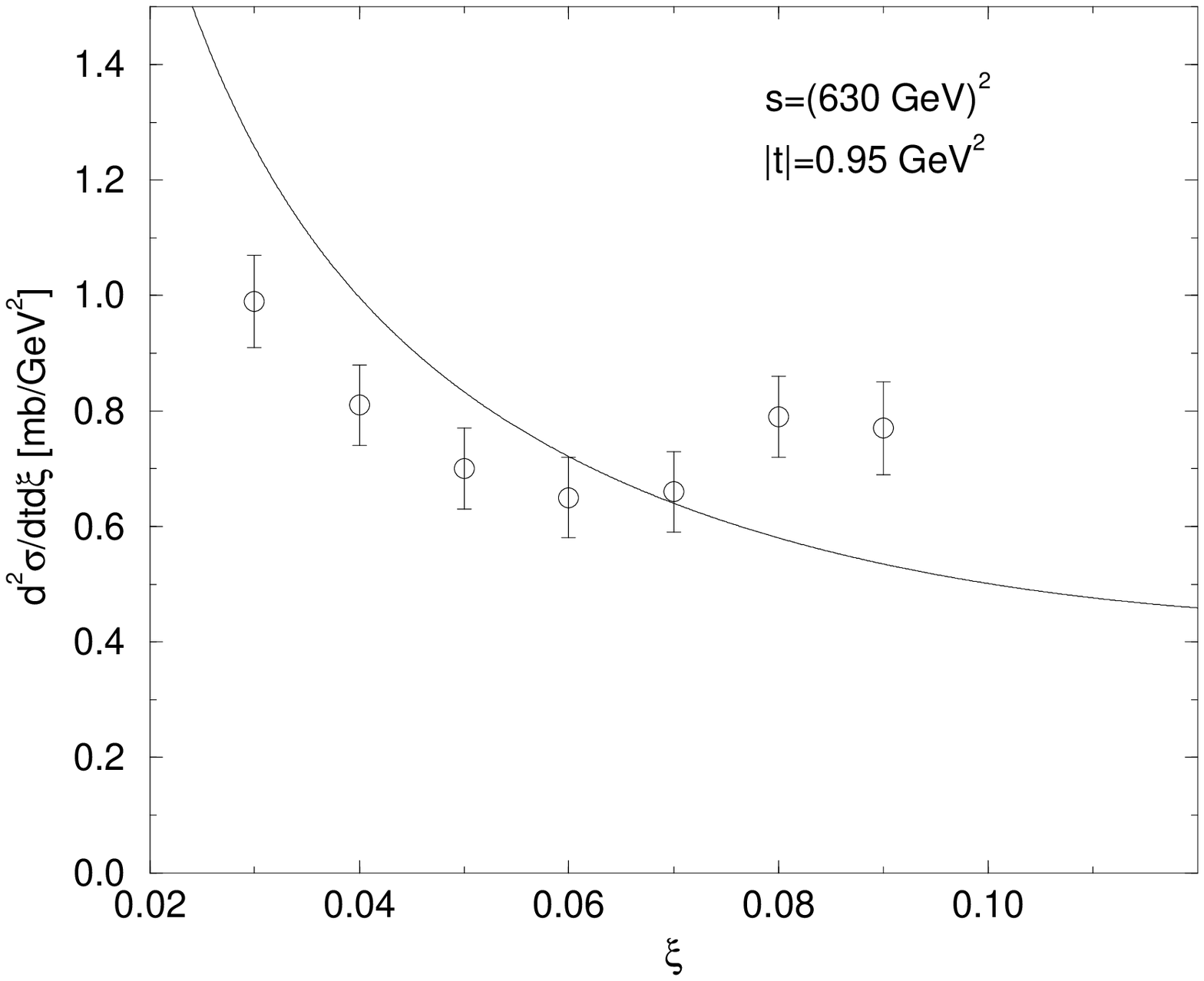}}}
\end{center}
\caption[]{Differential cross sections $d^2\sigma/d\xi\,dt$ [mb/GeV$^2$]
{\it vs} $\xi$ for $\sqrt s = 630$ GeV and $t=-0.95$ GeV$^2$. The 
solid curves represent the model with $\gamma = 2$ and with the other
parameters determined by the fit shown in Fig.~1. Data are 
from~\cite{UA8} and were not included in the fit.}
\end{figure}

Finally, we have considered the total single diffractive cross section 
$\sigma_{SD}$, for the process $p(\bar{p})+p\to p(\bar{p})+X$ as a function 
of $\sqrt{s}$. We have compared our model with the experimental data 
of~\cite{CDF,SCHA,ALB,ARM,BER} from the compilation given in~\cite{GM}, 
where some data have been corrected in order to obtain
the diffraction cross section for $\xi\leq 0.05$. In order to make the comparison,
we have numerically integrated our expression for $d^2\sigma/d\xi dt$, with
the parameters determined by the previous fit,
in the region $ 1.4/s \leq \xi \leq 0.05$ and  $t \leq 0$. In Fig.~4
we observe that the result of the integration, plotted as a function of $\sqrt s$,
is in good agreement with the experimental data over all
the range of values of $s$, including the Tevatron energies $\sqrt s =546$ and
1800 GeV. We must stress here that the choice $\gamma=2$ is essential: 
values of $\gamma$ lower than 2, but larger than 1/2 in order to satisfy 
$\sigma_{SD} < \sigma_T$, would give a too fast growth with $s$, whereas larger 
values of $\gamma$ would cause an undershooting of the data at large 
$s$. 

\begin{figure}[htb]
\begin{center}
{\parbox[t]{5cm}{\epsfysize 12cm \epsffile{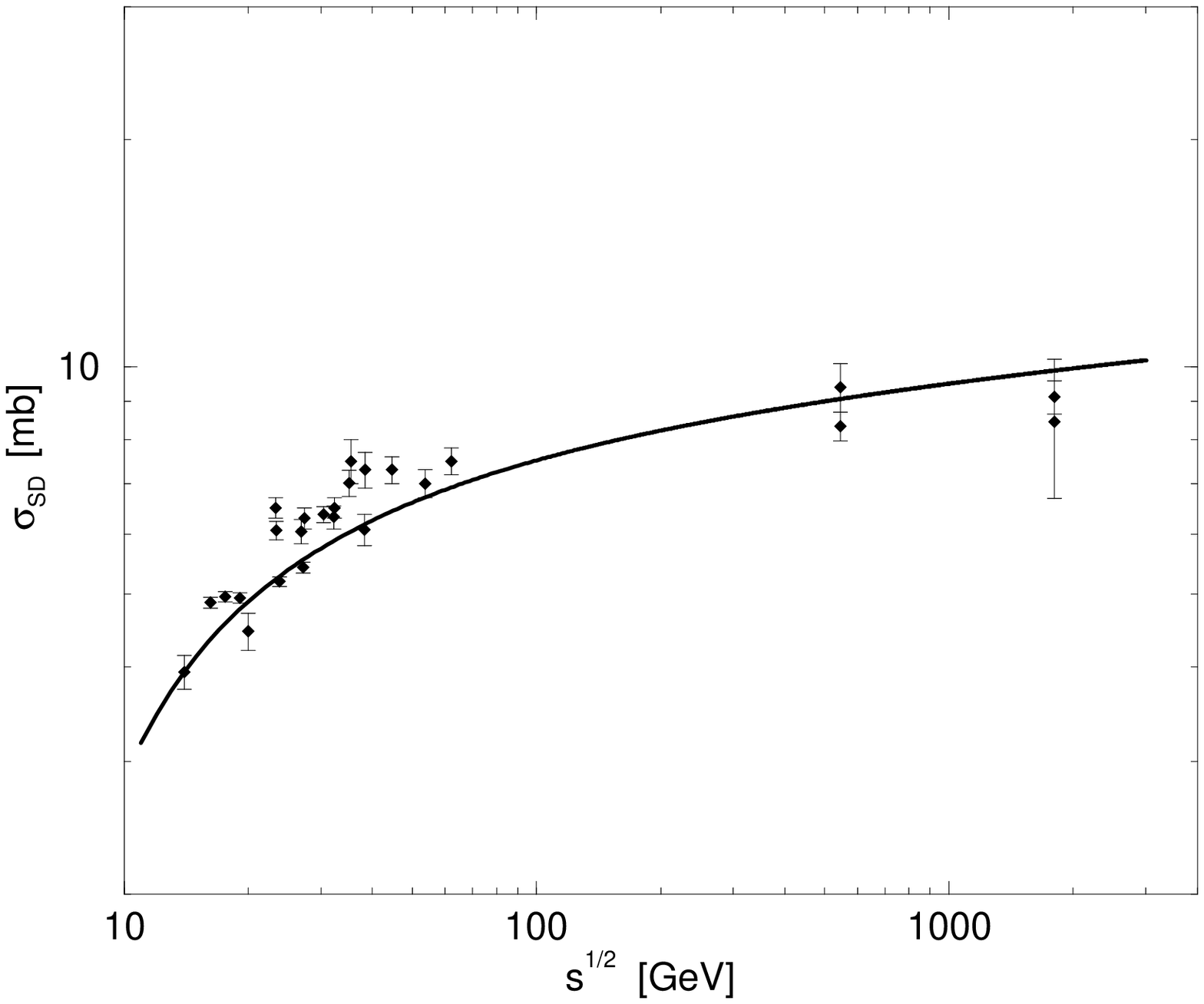}}}
\end{center}
\caption[]{Total single diffraction cross section $\sigma_{SD}$ {\it vs}
$\sqrt{s}$, compared with the prediction of the model. Data are
taken from the compilation of Ref.~\cite{GM}.}
\end{figure}

\section{Conclusions}

In this paper we have considered the proton diffraction dissociation
in the dipole Pomeron model. In this model the differential cross section
$d^2\sigma/d\xi dt$ can be written in the form given in Eq.~(\ref{z10}).
From the theoretical point of view, the result in Eq.~(\ref{z10}) assesses 
two important properties that seem to be required by the data~\cite{GM}.
First, the exact factorization, typical of the Regge pole model, is
lost in the dipole Pomeron approach. Second, for $t=0$ the
Pomeron and pion contributions are independent of $s$ and the
scaling with $M^2$ of $d^2\sigma/dM^2 dt |_{t=0}$ becomes exact if only
these terms are considered.
Moreover, we remark that this model respects the unitarity condition
without decoupling of the triple Pomeron vertex. The total
diffractive cross section rises as $\ln ( \ln s) $, i.e. slower than the total
$p-\bar{p}$ cross section that, in turn, satisfies the Froissart bound.

We notice that, in Eq.~(\ref{z10}), the triple Pomeron coupling and the 
Pomeron-proton cross section are tangled in the multiplicative constant 
$A$ together with an unknown scale factorized from the function $h(t)$.
Hence the fit of the experimental data cannot determine the aforesaid 
quantities but, at any rate, it represents an important test of the 
model.
Concerning the comparison with experimental data, we have found that 
this model gives a satisfactory fit to the experimental data for 
$d^2\sigma/d\xi dt$ with regards both to the $\xi-$ and $t-$dependence. 
Moreover, for a suitable choice of the parameter $\gamma$, it well reproduces
also the data for the total single diffractive cross section and allows
to predict a value of about 11 mb at the LHC energy $\sqrt s = 14$ TeV.

We stress that in our model the one-pion contribution, parametrized in
Eq.~(\ref{z8}), has been fixed from the beginning, differently from
Ref.~\cite{GM}, where a multiplicative constant has been considered in front of 
it as one of the two free parameters to be fitted.  As for the $f$ 
contribution, in our model it is well below the limit found by CDF~\cite{CDF}.
The discrepancies observed at large $\xi$ from the data of
Ref.~\cite{akim77} and of UA8~\cite{UA8} could arise from an 
underestimation of the contribution of the $\pi$ and from neglecting that
of the $\omega$. According to Ref.~\cite{UA8},  
the one-pion exchange contribution is only a small part of the total non-Pomeron 
exchange background.
Also the approximated treatment of the $f$ could be responsible for the 
disagreement at large $\xi$. 
What we need is a more rigorous method for  justifying the 
$t$-dependence of our parametrization and an extensive study of non-leading 
contributions. We feel that a deeper insight in these problems is
important for applications of the model to other processes.

\section{Acknowledgment}

One of us (L.L.J.) is grateful to the Dipartimento di Fisica dell'Universit\`a 
della Calabria and to the Istituto Nazionale di Fisica Nucleare - Sezione di Padova
e Gruppo Collegato di Cosenza for their warm hospitality and financial support.

\vfill\eject

\end{document}